\documentclass[twocolumn]{aastex62}
\usepackage{natbib,placeins,amsmath}

\newcommand{\msun}{\ifmmode M_{\odot}\else$M_{\odot}$\fi}
\newcommand{\rsun}{\ifmmode R_{\odot}\else$R_{\odot}$\fi}
\newcommand{\degrees}{\ifmmode^{\circ}\else$^{\circ}$\fi}
\newcommand{\amin}{\ifmmode^{\prime}\else$^{\prime}$\fi}
\newcommand{\asec}{\ifmmode^{\prime\prime}\else$^{\prime\prime}$\fi}
\newcommand{\cmmnt}[1]{\ignorespaces}

\submitjournal{ApJ}

\shorttitle{Fourier Domain Jerk Search}
\shortauthors{Andersen \& Ransom}

\begin{document}


\title{A Fourier Domain ``Jerk'' Search for Binary Pulsars}


\author[0000-0001-5908-3152]{Bridget C. Andersen}
\affil{University of Virginia, Department of Astronomy, P.O.~Box 400325, Charlottesville, VA 22904, USA}
\correspondingauthor{B.\,C.\,A.}
\email{bca9vh@virginia.edu}

\author[0000-0001-5799-9714]{Scott M.~Ransom}
\affil{National Radio Astronomy Observatory, 520 Edgemont Rd., Charlottesville, VA 22903, USA}
\affil{University of Virginia, Department of Astronomy, P.O.~Box 400325, Charlottesville, VA 22904, USA}

\begin{abstract}
While binary pulsar systems are fantastic laboratories for a wide array of astrophysics, they are particularly difficult to detect. The orbital motion of the pulsar changes its apparent spin frequency over the course of an observation, essentially ``smearing'' the response of the time series in the Fourier domain.  We review the Fourier domain acceleration search (FDAS), which uses a matched filtering algorithm to correct for this smearing by assuming constant acceleration for a small enough portion of the orbit. We discuss the theory and implementation of a Fourier domain ``jerk'' search, developed as part of the \textsc{PRESTO} software package, which extends the FDAS to account for a linearly changing acceleration, or constant orbital jerk, of the pulsar. We test the performance of our algorithm on archival Green Bank Telescope observations of the globular cluster Terzan~5, and show that while the jerk search has a significantly longer runtime, it improves search sensitivity to binaries when the observation duration is $5$ to $15\%$ of the orbital period. Finally, we present the jerk-search-enabled detection of Ter5am (PSR~J1748$-$2446am), a new highly-accelerated pulsar in a compact, eccentric, and relativistic orbit, with a likely pulsar mass of 1.649$^{+0.037}_{-0.11}$\,\msun.
\end{abstract}


\section{Introduction}
Binary pulsar systems produce extraordinary physics and astrophysics, yet detecting them can be a particularly difficult challenge. Isolated pulsars are detected by taking the fast Fourier transform (FFT) of an observational time series and searching through the resulting power spectrum. Since pulsars generally have narrow pulse profiles, resulting in many significant harmonics of the spin frequency in the power spectrum, those harmonics are summed to increase the significance of the final detection \citep[e.g.][]{lk12}.

However, in the case of binary pulsars, detection is complicated by the orbital motion of the pulsar. If the observation duration $T$ is longer than even a small fraction of the orbital period $P_{orb}$, Doppler shifting causes the apparent pulsar spin frequency to change with time. The Fourier power of the signal is ``smeared'' across neighboring frequency bins, leading to a decrease in the significance of the detection. Higher harmonics of the pulsar signal are impacted more than the fundamental, as Doppler smearing increases proportionally with harmonic number.

Acceleration searches account for this Doppler smearing by assuming that, over a small enough fraction of the orbit ($T \lesssim P_{orb}/10$), the pulsar's acceleration is approximately constant, and therefore its measured spin frequency drifts linearly in time \citep[e.g.][]{jk91}. Blind acceleration searches iterate over many constant-acceleration values, completing data manipulations in the time or Fourier domain to recover signal power that the orbital motion has smeared over several frequency bins.

The higher search dimensionality and computational expense of acceleration searches has precluded their use in wide pulsar surveys until this past decade. However, for high-priority targeted searches, such as those towards supernova remnants \citep[e.g.][]{mk84}, globular clusters \citep[e.g.][]{clf+00}, and {\em Fermi} unassociated sources \citep[e.g.][]{rab+12}, acceleration searches have uncovered more than $100$ binary pulsars. The vast majority were found using a Fourier domain implementation of the acceleration search \citep{rgh+01,rem02}, which has recently become known as the Fourier Domain Acceleration Search \citep[FDAS;][]{dat+18}\footnote{Other algorithms have been developed to detect binary pulsars under different observational conditions, including the Dynamic Power Spectrum when $T \sim P_{orb}$ \citep{cha03} and phase modulation searches when $T \gtrsim 2P_{orb}$ \citep{rce03}.}.

While the FDAS is effective when $T \lesssim P_{orb}/10$, search sensitivity to fainter pulsars improves with longer observing duration as $\sqrt{T}$. Acceleration searches are therefore limited to discovering only the brightest binary pulsars. To enable the detection of fainter and more accelerated systems, we can add another level of approximation to the acceleration search by assuming that the next derivative of orbital motion, the ``jerk,'' is constant. Under this assumption, the measured spin frequency changes quadratically with time and the spin phase changes cubically.

Jerk searches allow for longer observation durations covering larger portions of the pulsar orbit than acceleration searches, while still retaining significant power in the Fourier domain \citep{blw13}. Therefore, although adding the jerk dimension substantially increases the runtime of the search, it also allows us to probe for systems previously missed with acceleration searches due to residual Doppler smearing. This capability will be especially useful in targeted searches for more exotic systems like Galactic Center pulsars, pulsar-black hole binaries, and the most compact and relativistic double neutron star systems.

We have developed and tested a Fourier domain jerk search as part of the \textsc{PRESTO}\footnote{\url{https://github.com/scottransom/presto}} software package \citep{ran01}. This paper briefly describes the details and performance of our implementation, and presents the detection of a new pulsar in the globular cluster Terzan~5 found using our jerk search.

\section{Jerk Search Theory and Implementation} \label{TheoryAndImplement}
Our jerk search functions as a fairly straightforward extension to the original FDAS implemented in \textsc{PRESTO}'s \texttt{accelsearch} program. The mathematics and methodology of this FDAS are described in detail in \cite{rem02} and \cite{dat+18}. In the following section, we provide a brief summary.

\subsection{Acceleration Search Review}
In short, the FDAS is a matched filtering algorithm that corrects for Doppler smearing by assuming that the pulsar's acceleration, $\alpha$, is roughly constant over the course of the observation. Under this assumption, each harmonic of the pulsar signal would experience a constant frequency derivative, $\dot f$, according to the relation
\begin{equation}
	\alpha = \frac{\dot{fc}}{hf} = \frac{zc}{hfT^2},
\end{equation}
where $c$ is the speed of light, $T$ is the observation duration, $f$ is the frequency of the fundamental, and $h$ is the harmonic number where $h=1$ represents the fundamental. In this equation we also introduce the Fourier frequency derivative, $z$. Fourier frequencies (i.e.~wavenumbers or FFT bin numbers) are defined as $r = fT$, and so $z = \dot r = \dot fT^2$ corresponds to the number of Fourier frequency bins that the signal drifts through over the course of the observation. Similarly, $w = \dot z = \ddot fT^3$ corresponds to the Fourier jerk, or the number of frequency derivative bins that the signal drifts through over the course of the observation. We use this $r, z, w$ coordinate system throughout our acceleration and jerk search code, as it is computationally and intuitively advantageous when dealing with the properties of discrete Fourier transforms.  When using these Fourier-based units defined over the whole observation, it is also convenient to switch to a normalized time coordinate, $u$, which represents the fraction of the observation complete at any given instant (such that $0 \leq u \leq 1$). 

Using the Convolution Theorem, the acceleration search correlates template Fourier domain amplitude and phase responses for a number of trial accelerations with the complex Fourier amplitudes from an FFT of the original input time series. The correlations are completed as a series of short FFTs using a computationally efficient ``overlap-and-save'' technique \citep[see \S3 of][]{dat+18}.

\begin{figure*}
\begin{center}
\includegraphics[width=7in]{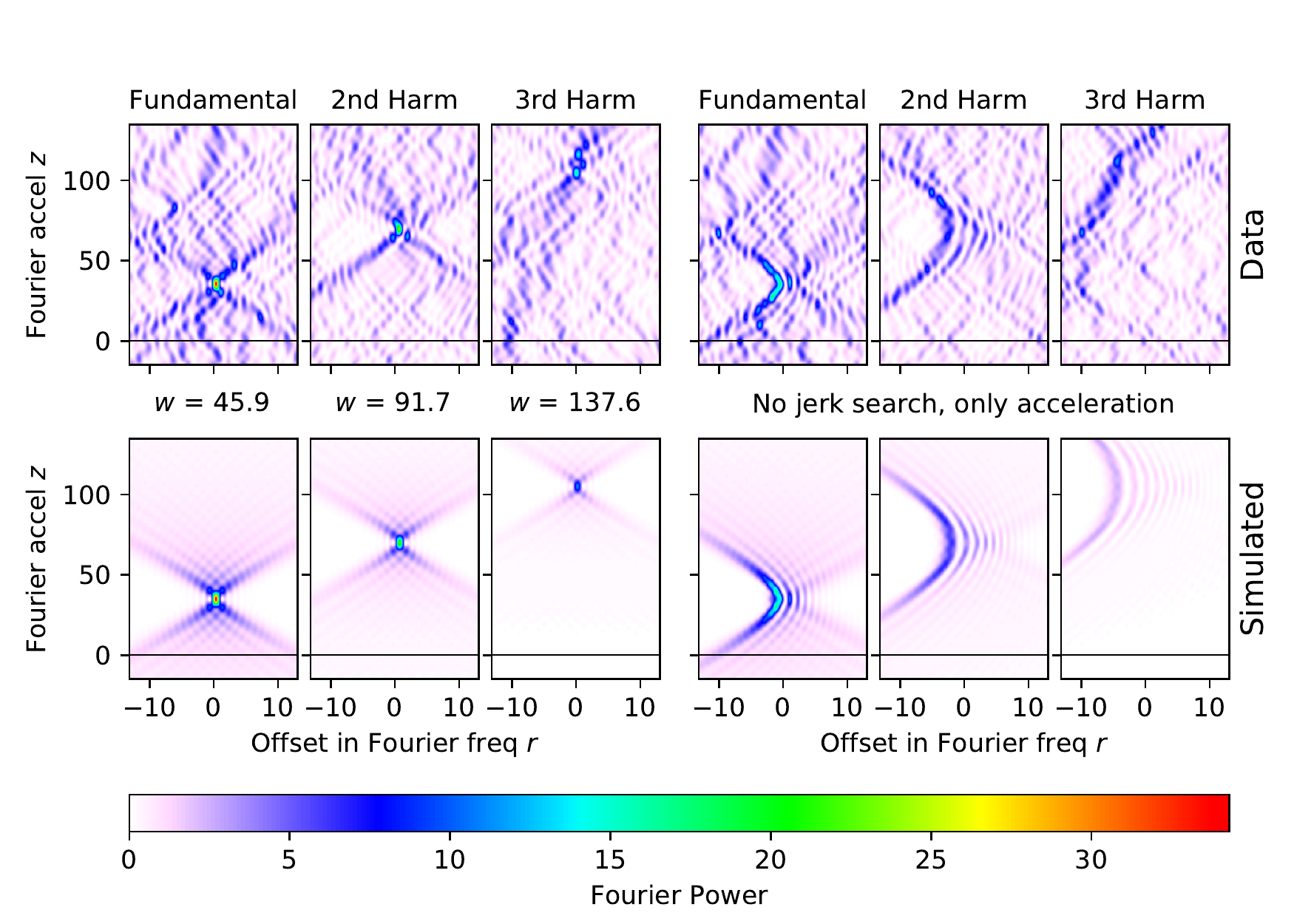}
\caption{Detection of the first three harmonics of the pulsar J1748$-$2446M (aka~Ter5M) from the exact same portion of the GBT observation where we found the new pulsar, J1748$-$2446am. Each image shows powers from slices in the $f$-$\dot{f}$ plane (i.e.~Fourier $r$ vs.~Fourier $z$) at specific values of the Fourier jerk $w$. The top and bottom rows show the data or a simulated noiseless response after properly correcting for the jerk of the pulsar during the observation (left three columns) or using only an acceleration search (i.e.~with no correction for jerk, right three columns). The detection significance is much higher using the jerk search as more harmonics are detected and each of those harmonics has more power. A completely un-accelerated search would correspond to searching the $z = 0$ line in the acceleration-only search. The simulated signals also effectively show the shapes of some of the Fourier domain search templates for acceleration (left; when jerk is zero or fully corrected) or jerk searches (right; when nonzero jerk leads to asymmetric templates).} \label{fig:FFdotplanes}
\end{center}
\end{figure*}

When stacked according to the trial $\dot f$ value, the resulting power spectra from these correlations form a $2$D plane in Fourier frequency vs.~frequency derivative space. Fig.~\ref{fig:FFdotplanes} shows examples of such $f$-$\dot{f}$ planes for both simulated and actual pulsar signals. Once the $f$-$\dot{f}$ plane is constructed, the algorithm searches through it and identifies candidate pulsar signatures according to a pre-calculated power threshold. The search also incoherently sums the powers from a number of harmonics onto the fundamental to increase the probability of detecting narrow pulse profiles.

For an acceleration search, the template responses can be calculated analytically \citep[see \S4.2.2 of][for a complete mathematical derivation]{rem02}. With a constant acceleration, we can approximate a harmonic of the pulsar signal as a sinusoid with a quadratically varying phase. Taking the Fourier transform of this signal yields an analytic expression for the template, composed of Fresnel integrals. The simulated signal in the bottom-left panels of Fig.~\ref{fig:FFdotplanes}, which show the response of an accelerated (but jerk-corrected) pulsar across three harmonics, also effectively demonstrates the distinctive ``X'' shape that these templates form when displayed in the $f$-$\dot{f}$ plane. In the bottom-right panels, the $z=0$ lines show the powers of the initial FFT of the time series, without any sort of acceleration or jerk correction.

\subsection{Jerk Search}\label{theory}
Just as the acceleration search is rooted in the approximation of constant acceleration via a linearly varying spin frequency, our jerk search approximates a constant jerk with linearly varying acceleration, or a quadratically varying spin frequency. A constant jerk corresponds to a constant second time-derivative of the frequency,
\begin{equation}
	\dot{\alpha} = \frac{\ddot{f}c}{hf} = \frac{wc}{hfT^3}, \label{eq:dimoudi2equiv}
\end{equation}
where $\dot{\alpha}$ is the jerk, $\ddot{f}$ is the second time-derivative of the frequency, and $w$ is the Fourier jerk.

The actual mechanics of the jerk search are very similar to what we have just described for the acceleration search. We use the same overall process of overlap-and-save matched filtering, threshold searching, and harmonic summing, except instead of generating an $f$-$\dot{f}$ plane of powers, we arrange the correlation results in an $f$-$\dot{f}$-$\ddot{f}$ (or $r$-$z$-$w$) volume which we then search for candidates.

Generating this $r$-$z$-$w$ volume requires us to calculate different template responses for correlation. Under the assumption of constant jerk, we approximate a harmonic of the pulsar signal $n(u)$ as a sinusoid with a cubically varying phase
\begin{equation}
	n(u) = a\cos\left[2\pi\left(r_0u + \frac{1}{2} z_0u^2 + \frac{1}{6}wu^3\right) + \phi\right],  \label{eq:jerksignal2}
\end{equation}
where $a$ is the amplitude of the signal, $r_0$ and $z_0$ are the Fourier frequency and frequency derivative at the start of the observation, and $\phi$ is a starting phase. To make our jerk search easier to implement, we want our collection of templates to be as compact in $r$, $z$, $w$ space as possible around the template origin. To accomplish this, we can define our initial Fourier frequency and frequency derivative values, $r_0$ and $z_0$, in terms of the average frequency and average frequency derivative over the course of the observation, $\bar{r}$ and $\bar{z}$,
\begin{align*}
	r_0 = \bar{r} - \frac{\bar{z}}{2} + \frac{w}{12}\ \textrm{and}\ z_0 = \bar{z} - \frac{w}{2}.
\end{align*}
These relations come from averaging $r(u) = d\Phi/du$ and $z(u) = d^2\Phi/du^2$ from $u=0$ to $u=1$, where $\Phi$ is the phase of the cosine in Eqn.~\ref{eq:jerksignal2}.

Expanding Eqn.~\ref{eq:jerksignal2} into complex exponentials gives
\begin{align}
	n(u) &= \frac{a}{2}\left\{e^{2\pi i \left[(\bar{r} - \frac{\bar{z}}{2} + \frac{w}{12})u + \frac{1}{2} (\bar{z} - \frac{w}{2})u^2 + \frac{1}{6}wu^3\right]} e^{i\phi} \right. \nonumber \\ 
    &+ \left. e^{-2\pi i \left[(\bar{r} - \frac{\bar{z}}{2} + \frac{w}{12})u + \frac{1}{2} (\bar{z} - \frac{w}{2})u^2 + \frac{1}{6}wu^3\right]} e^{-i\phi} \right\}.\label{eq:expjerk}
\end{align}
The Fourier response of the jerked signal, at frequency $r$, is the Fourier transform of $n(u)$,
\begin{align}
	A_r &= N \int_0^1 n(u) e^{-2\pi iru} du \nonumber \\
    &= \frac{aN}{2} e^{i\phi}\int_{0}^{1} e^{2\pi i \left[(\Delta r - \frac{\bar{z}}{2} + \frac{w}{12})u + \frac{1}{2} (\bar{z} - \frac{w}{2})u^2 + \frac{1}{6}wu^3\right]}\ du, \label{eq:jerkintegral} 
\end{align}
where $N$ is the number of samples in our original time series, $\Delta r = \bar{r} - r$, and the second term in Eqn.~\ref{eq:expjerk} averages to zero.  The correlation templates for the search are the complex conjugates of Eqn.~\ref{eq:jerkintegral} over a range of $\Delta r$.

Unlike the template integral for an acceleration search, Eqn.~\ref{eq:jerkintegral} has no analytic solution. However, as in the acceleration case, the template shape is independent of the absolute value of $r_0$ or $\bar{r}$. This independence allows us to simulate the portion of the binary orbit described by Eqn.~\ref{eq:jerksignal2} efficiently in a time series of only a few hundred or a few thousand points, depending on $\bar{z}$ and $w$. Taking an FFT of the time series then computes the Fourier integral numerically. The simulated signals in the bottom-right panel of Fig.~\ref{fig:FFdotplanes} show the uncorrected response of the first three harmonics of a jerked pulsar when displayed in the $f$-$\dot{f}$ plane (located at $w = 0$ in the $f$-$\dot{f}$-$\ddot{f}$ volume).

Once portions of the $f$-$\dot{f}$-$\ddot{f}$ volume are computed, they are typically converted to normalized powers, and then either incoherently added to other portions of the $f$-$\dot{f}$-$\ddot{f}$ volume at signal harmonics, or searched directly for significant signals \citep[see e.g.~][]{rem02}. As per a normal Fourier search, the power-normalized incoherent harmonic sums are $\chi^2$-distributed with $2m$ degrees of freedom, where $m$ is the number of summed harmonics.  Determining the overall significance of a signal is beyond the scope of this paper, however, for well-behaved data (i.e.~a relatively uniformly sampled and white-noise-dominated input time series), the approximate number of independent Fourier bins searched must be accounted for (i.e.~a search trials factor).  For a jerk search, the approximate number of independent trials for a single choice of the number of harmonics to sum $m$, is 
\begin{equation} \label{eq:Ntrials}
	N_{\text{trials}} \sim \frac{N_r}{m} \left(\frac{N_{z}}{6.95}\right)\left(\frac{N_{w}}{44.2}\right),
\end{equation}
where $N_r$, $N_z$, and $N_w$ are the numbers of Fourier frequency bins, $z$ bins, and $w$ bins searched for the highest harmonic summed.  The numerical values $6.95$ and $44.2$ are the Fourier widths at half-power of the Fourier response in the $z$ and $w$ directions, respectively.

When conducting searches, in order to help prevent ``scalloping'' of a signal's power due to the finite grid of computed $r$, $z$, and $w$ points, \texttt{accelsearch} oversamples the volume in each of those directions. The grid spacing used is 0.5 (i.e.~interbinning or Fourier interpolation), 2, and 20, for $r$, $z$, and $w$, respectively, given the measured half-widths of the peaks in each of those orthogonal directions (see above). To determine what range of $z$ and $w$ values to search, we simulated thousands of realistic pulsar binaries and ``observed'' them with integrations of tens of minutes to hours. Most systems are detected with $|z| < 200$ and $|w| < 600$. For larger values of $z$ or $w$, given realistic orbits with stellar mass companions, regardless of $T/P_{orb}$, the constant acceleration or constant jerk assumption breaks down and we lose sensitivity.

\begin{deluxetable}{lc}
\tablecaption{PSR J1748$-$2446am\label{tab1}}
\tablewidth{0pt}
\startdata
\\
\multicolumn{2}{c}{Timing Parameters}\\
Right Ascension (RA, J2000) \dotfill & $17^{\rm h}\;48^{\rm m}\;04\fs8235(2)$\\
Declination     (DEC, J2000) \dotfill & $-24\degrees\;46\amin\;47\farcs21(9)$\\
Pulsar Period (ms) \dotfill & 2.933819877244(2)\\
Pulsar Frequency (Hz) \dotfill & 340.8525546358(2)\\
Frequency Derivative (Hz\,s$^{-1}$) \dotfill & 1.5893(4)$\times$10$^{-14}$\\
Reference Epoch (MJD) \dotfill & 53700\\
Dispersion Measure (pc cm$^{-3}$) \dotfill & 238.193(3)\\
Orbital Period (days) \dotfill & 0.80010926(2)\\
Projected Semi-Major Axis (lt-s)  \dotfill & 0.937815(5)\\
Orbital Eccentricity  \dotfill & 0.204736(9)\\
Epoch of Periastron (MJD) \dotfill & 53700.440278(6)\\
Longitude of Periastron, $\omega$ (deg) \dotfill & 337.365(2)\\
Time Derivative of $\omega$, $\dot\omega$ (deg\,yr$^{-1}$) \dotfill & 0.454(4)\\
Span of Timing Data (MJD) \dotfill & 53193$-$54195\\
Number of TOAs  \dotfill & 217\\
RMS TOA Residual ($\mu$s) \dotfill & 38.5\\
\multicolumn{2}{c}{Derived Parameters}\\
Mass Function (\msun) \dotfill & 0.00138336(2)\\
Total System Mass (\msun) \dotfill & 1.85(2)\\
Min Companion Mass (\msun) \dotfill & $\geq$\,0.15\\
Companion Mass (\msun) \dotfill & 0.194 ($+$0.11, $-$0.023)\\
Pulsar Mass (\msun) \dotfill & 1.649 ($+$0.037, $-$0.11)\\
Flux Density at 2\,GHz (mJy) \dotfill & $\sim$0.015\\
\enddata
\tablecomments{Numbers in parentheses represent 1-$\sigma$
uncertainties in the last digit.  The timing solution was
determined using {\tt tempo} with the DE436 Solar System Ephemeris and
the DD binary model. The time system used is Barycentric Dynamical
Time (TDB).  The minimum companion mass was calculated assuming a
pulsar mass of 1.4\,\msun.  The total system mass and 68\% central
confidence ranges on the masses of the pulsar and its companion were
determined assuming that $\dot\omega$ is due completely to general
relativity, and a random orbital inclination (i.e.~probability density
is uniform in $\cos i$).}
\end{deluxetable}

The slices through the jerk volume displayed in Fig.~\ref{fig:FFdotplanes} illuminate other salient properties of the jerk search that are worth noting. First off, we can see that when the jerk volume is sliced through at the $w$ of the pulsar signal (in this case $45.9$), the response of the signal in the resulting $f$-$\dot{f}$ plane is the characteristic ``X'' shape that we expect from a standard acceleration search \citep[see][]{rem02}, indicating that the smearing due to jerk has been mitigated\footnote{For an animated projection through the $r$-$z$-$w$ volume for a sinusoid, see \url{http://www.cv.nrao.edu/~sransom/ffdot_wrange.gif}}. Another important feature to note is that the higher harmonics of the signal are essentially ``jerked'' out of significance in the acceleration search. This is just as predicted by Eqn.~\ref{eq:dimoudi2equiv}, which shows that as the harmonic number (and therefore frequency, $f$) increases and the jerk remains constant, the Fourier $w$ of the signal must increase proportionally. Thus, higher harmonics are more heavily affected by jerk. This effect is clearly illustrated in the left panels of Fig.~\ref{fig:FFdotplanes}. As the frequency doubles from the fundamental to the second harmonic, the Fourier $w$ also doubles from $45.9$ to $91.7$. As a result, the detection significance is much higher using the jerk search as each of the harmonics has more power to contribute during harmonic summing.

\section{Performance} \label{Performance}
While developing and testing the algorithm and its implementation in \texttt{accelsearch}, we repeatedly searched two archival Robert C.~Byrd Green Bank Telescope (GBT) observations of the globular cluster Terzan~5 taken 2005 May 15 and 2008 September 12. These data, described in \S\ref{Detection} and in \citet{rhs+05}, were dedispersed at dispersion measures (DMs) of 238.00 and 238.72\,pc\,cm$^{-3}$, respectively.  Each observation contains numerous binary MSPs with relatively short orbital periods ($P_b \lesssim 1$\,day) that are detectable with significant accelerations and jerks over integrations between 10\,min to the $\sim$7\,hr durations of the observations. 

As an example of how detection significances can vary for real pulsars, we describe how five different Terzan~5 MSPs were detected in a single search of a 4096-s segment of the 2005 data, using \texttt{accelsearch -numharm 4 -zmax 300} and with and without \texttt{-wmax 900}. Each of the reported detection significances are in $\sigma$ (i.e.~equivalent gaussian significance) after correcting for the approximate number of independent trials searched (see \S\ref{theory}), which is $\sim$41 times larger for the jerk searches than the acceleration searches, using these parameters.

For the isolated MSP Ter5L, the acceleration-only search detected the pulsar with a significance of 7.4\,$\sigma$, compared to 6.6\,$\sigma$ for the jerk search.  Similarly, the weakly accelerating long-period binary Ter5E was detected at 8.9 and 8.2\,$\sigma$, respectively, although the total summed power was larger in the jerk search.  These results show, as expected, that you pay a penalty with a jerk search for weakly- or un-accelerated pulsars due to the larger phase space searched.  The situation was different for the compact binaries Ter5I, M, and N.  Ter5I and Ter5M were detected with 1.2 and 3.0 extra sigma in the jerk searches, although Ter5N lost 0.9\,$\sigma$ since $w\sim0$ during that portion of the pulsar's orbit.

These test searches, as well as the thorough analysis performed by \citet{blw13}, show that jerk searches can improve the sensitivity to highly accelerating pulsars with $T \sim 0.05-0.15 P_{orb}$ by a significant amount. The penalty is a slightly reduced sensitivity to weakly- or un-accelerated pulsars due to the additional independent trials searched, as well as a substantially longer run time that, in this case, increased by a factor of almost 80. In general, the runtime is roughly linear with the number of trials according to Eqn.~\ref{eq:Ntrials}.

\section{Detection of a New Pulsar: Ter5am} \label{Detection}
While searching eleven overlapping segments of duration 4096\,s from the 2008 observation, with \texttt{-zmax 100} and \texttt{-wmax 500}, we detected a highly accelerated new pulsar with a spin period of 2.93\,ms, $z=-21$, and $w=-15$ in one segment, using a four-harmonic sum\footnote{We also detected 9 other known binary pulsars showing accelerations and jerks: Ter5A, E, I, J, M, N, V, ae, and ai. For a full list of pulsars in the cluster, see: \url{http://www.naic.edu/~pfreire/GCpsr.html}}. A similar search, using acceleration but no jerk, did not detect the pulsar, mostly because the higher harmonics had been ``jerked'' out of significance.

After determining a better DM for the pulsar, we searched other archival observations and detected it several additional times, allowing us to solve the compact and eccentric orbit, and eventually determine a fully-coherent timing solution. Here we present the timing from the first $\sim$1000\, days of Terzan~5 observations made with the GBT Pulsar Spigot \citep{kel+05}. Most of the observations were taken at 2\,GHz with a usable bandwidth of $\sim$600\,MHz (out of 800\,MHz total), a sample time of 81.92\,$\mu$s, and 2048\,frequency channels \citep[see][for details]{rhs+05,hrs+06}.

The timing solution for PSR~J1748$-$2446am includes a strong detection of the advance of periastron (i.e.~$\dot \omega$).  If entirely relativistic, which is likely given a white dwarf companion, the total mass of the system is 1.85$\pm$0.02\,\msun, and the component masses can be constrained with the mass function by assuming random inclinations (see Table~\ref{tab1}), in which case, the pulsar mass is 1.649$^{+0.037}_{-0.11}$\,\msun. A longer-duration timing solution, comprising $\sim$14\,yrs of Spigot and coherently-dedispersed GUPPI data will be presented elsewhere (Ransom et al., in prep.).

\section{Conclusions} \label{Conclusions}
In this paper we have presented the implementation and performance of a new Fourier domain jerk search, which extends the \textsc{PRESTO} FDAS to account for linearly varying acceleration (i.e.~constant jerk). Consistent with the analysis conducted by \cite{blw13}, we find that our algorithm can significantly improve search sensitivity to pulsar binaries where $T \sim 0.05 - 0.15 P_{orb}$. Trade-offs of the algorithm include a significantly longer runtime and decreased sensitivity to un-accelerated systems due to the increased independent trials factor. While testing our algorithm on GBT observations of the globular cluster Terzan~5, we discovered Ter5am, an interesting compact eccentric binary that already shows relativistic periastron advance. Other relativistic effects (such as relativistic $\gamma$) will likely become detectable in coming years, allowing us to obtain precise mass measurements for both objects in the system.

Looking to the future, this jerk search technique will be especially useful in high-profile targeted searches for exotic systems such as Galactic Center and globular cluster pulsars, pulsar-black hole binaries, and the most compact and relativistic double neutron star systems. Since each of the template correlations in the $r$-$z$-$w$ volume are independent of each other, this technique also has parallelization potential. While the current implementation makes use of \textsc{OpenMP}, a straightforward extension of the \citet{dat+18} methodology would allow us to implement the jerk search on GPUs. This could significantly reduce the runtime cost of the algorithm, opening it up to wider use and less focused searches.

\acknowledgments
The National Radio Astronomy Observatory is a facility of the National Science Foundation operated under cooperative agreement by Associated Universities, Inc.  The Green Bank Observatory is a facility of the National Science Foundation operated under cooperative agreement by Associated Universities, Inc.  SMR is a CIFAR Senior Fellow and is also supported by the NSF Physics Frontiers Center award number 1430284.  We thank Jason Hessels, Ingrid Stairs, and Paulo Freire for their help with GBT observations of Terzan~5 over the years, and Thankful Cromartie for a careful reading of the manuscript.

\software{PRESTO \citep{ran01,presto}, Tempo \citep{tempo}}
\facility{GBT}


\begin{thebibliography}{}
\expandafter\ifx\csname natexlab\endcsname\relax\def\natexlab#1{#1}\fi

\bibitem[{{Bagchi} {et~al.}(2013){Bagchi}, {Lorimer}, \& {Wolfe}}]{blw13}
{Bagchi}, M., {Lorimer}, D.~R., \& {Wolfe}, S. 2013, \mnras, 432, 1303

\bibitem[{{Camilo} {et~al.}(2000){Camilo}, {Lorimer}, {Freire}, {Lyne}, \&
  {Manchester}}]{clf+00}
{Camilo}, F., {Lorimer}, D.~R., {Freire}, P., {Lyne}, A.~G., \& {Manchester},
  R.~N. 2000, \apj, 535, 975

\bibitem[{{Chandler}(2003)}]{cha03}
{Chandler}, A.~M. 2003, PhD thesis, California Institute of Technology

\bibitem[{{Dimoudi} {et~al.}(2018){Dimoudi}, {Adamek}, {Thiagaraj}, {Ransom},
  {Karastergiou}, \& {Armour}}]{dat+18}
{Dimoudi}, S., {Adamek}, K., {Thiagaraj}, P., {et~al.} 2018, ArXiv e-prints,
  arXiv:1804.05335, {A}pJ, in press

\bibitem[{{Hessels} {et~al.}(2006){Hessels}, {Ransom}, {Stairs}, {Freire},
  {Kaspi}, \& {Camilo}}]{hrs+06}
{Hessels}, J.~W.~T., {Ransom}, S.~M., {Stairs}, I.~H., {et~al.} 2006, Science,
  311, 1901

\bibitem[{{Johnston} \& {Kulkarni}(1991)}]{jk91}
{Johnston}, H.~M., \& {Kulkarni}, S.~R. 1991, \apj, 368, 504

\bibitem[{{Kaplan} {et~al.}(2005){Kaplan}, {Escoffier}, {Lacasse}, {O'Neil},
  {Ford}, {Ransom}, {Anderson}, {Cordes}, {Lazio}, \& {Kulkarni}}]{kel+05}
{Kaplan}, D.~L., {Escoffier}, R.~P., {Lacasse}, R.~J., {et~al.} 2005, \pasp,
  117, 643

\bibitem[{{Lorimer} \& {Kramer}(2012)}]{lk12}
{Lorimer}, D.~R., \& {Kramer}, M. 2012, {Handbook of Pulsar Astronomy}

\bibitem[{{Middleditch} \& {Kristian}(1984)}]{mk84}
{Middleditch}, J., \& {Kristian}, J. 1984, \apj, 279, 157

\bibitem[{{Nice} {et~al.}(2015){Nice}, {Demorest}, {Stairs}, {Manchester},
  {Taylor}, {Peters}, {Weisberg}, {Irwin}, {Wex}, \& {Huang}}]{tempo}
{Nice}, D., {Demorest}, P., {Stairs}, I., {et~al.} 2015, {Tempo: Pulsar timing
  data analysis}, Astrophysics Source Code Library, ascl:1509.002

\bibitem[{{Ransom}(2011)}]{presto}
{Ransom}, S. 2011, {PRESTO: PulsaR Exploration and Search TOolkit},
  Astrophysics Source Code Library, ascl:1107.017

\bibitem[{{Ransom}(2001)}]{ran01}
{Ransom}, S.~M. 2001, PhD thesis, Harvard University

\bibitem[{{Ransom} {et~al.}(2003){Ransom}, {Cordes}, \& {Eikenberry}}]{rce03}
{Ransom}, S.~M., {Cordes}, J.~M., \& {Eikenberry}, S.~S. 2003, \apj, 589, 911

\bibitem[{{Ransom} {et~al.}(2002){Ransom}, {Eikenberry}, \&
  {Middleditch}}]{rem02}
{Ransom}, S.~M., {Eikenberry}, S.~S., \& {Middleditch}, J. 2002, \aj, 124, 1788

\bibitem[{{Ransom} {et~al.}(2001){Ransom}, {Greenhill}, {Herrnstein},
  {Manchester}, {Camilo}, {Eikenberry}, \& {Lyne}}]{rgh+01}
{Ransom}, S.~M., {Greenhill}, L.~J., {Herrnstein}, J.~R., {et~al.} 2001, \apj,
  546, L25

\bibitem[{{Ransom} {et~al.}(2005){Ransom}, {Hessels}, {Stairs}, {Freire},
  {Camilo}, {Kaspi}, \& {Kaplan}}]{rhs+05}
{Ransom}, S.~M., {Hessels}, J.~W.~T., {Stairs}, I.~H., {et~al.} 2005, Science,
  307, 892

\bibitem[{{Ray} {et~al.}(2012){Ray}, {Abdo}, {Parent}, {Bhattacharya},
  {Bhattacharyya}, {Camilo}, {Cognard}, {Theureau}, {Ferrara}, {Harding},
  {Thompson}, {Freire}, {Guillemot}, {Gupta}, {Roy}, {Hessels}, {Johnston},
  {Keith}, {Shannon}, {Kerr}, {Michelson}, {Romani}, {Kramer}, {McLaughlin},
  {Ransom}, {Roberts}, {Saz Parkinson}, {Ziegler}, {Smith}, {Stappers},
  {Weltevrede}, \& {Wood}}]{rab+12}
{Ray}, P.~S., {Abdo}, A.~A., {Parent}, D., {et~al.} 2012, ArXiv e-prints,
  arXiv:1205.3089

\end{thebibliography}
\end{document}